# PrivBasis: Frequent Itemset Mining with Differential Privacy

Ninghui Li, Wahbeh Qardaji, Dong Su, Jianneng Cao
Purdue University
305 N. University Street,
West Lafayette, IN 47907, USA
{ninghui, wqardaji, su17, cao72}@cs.purdue.edu

## ABSTRACT

The discovery of frequent itemsets can serve valuable economic and research purposes. Releasing discovered frequent itemsets, however, presents privacy challenges. In this paper, we study the problem of how to perform frequent itemset mining on transaction databases while satisfying differential privacy. We propose an approach, called PrivBasis, which leverages a novel notion called basis sets. A $\theta$-basis set has the property that any itemset with frequency higher than $\theta$ is a subset of some basis. We introduce algorithms for privately constructing a basis set and then using it to find the most frequent itemsets. Experiments show that our approach greatly outperforms the current state of the art.

## 1. INTRODUCTION

Frequent itemset mining is a well recognized data mining problem. The discovery of frequent itemsets can serve valuable economic and research purposes, e.g., mining association rules [5], predicting user behavior [3], and finding correlations [11]. Publishing frequent itemsets, however, may reveal the information of individual transactions, compromising the privacy of them.

In this paper, we study the problem of how to perform frequent itemset mining (FIM) on transaction databases while satisfying differential privacy. Differential privacy [17] is an appealing privacy notion which provides worst-case privacy guarantees. In recent years, it has become the de facto standard notion of privacy for research in private data analysis. The key challenge in private FIM is that the dimensionality of transactional datasets is very high. While effective techniques for differentially private data publishing have been developed for low-dimensional datasets (e.g., [23, 33]), these techniques do not apply to high-dimensional data. In fact, even for the weaker privacy notion of $k$-anonymity, the curse of high dimensionality effect is well known [4].

Our work is inspired by Bhaskar et al.'s KDD10 paper [8], in which they propose an approach to privately publish top $k$ frequent itemsets and their frequencies. Their approach first selects $k$ itemsets from the set of all itemsets that include at most $m$ items, and then adds noise to the frequencies of these selected itemsets. This approach works reasonably well for small $k$ values; however for larger values of $k$, the accuracy is poor. The main reason is that for larger $k$ values, one has to set the size limit $m$ to be larger (e.g., 3, 4, or higher). This results in a very large candidate set from which the algorithm must select the top $k$, making the selection inaccurate.

In this paper we propose a novel approach that avoids the selection of top $k$ itemsets from a very large candidate set. More specifically, we introduce the notion of basis sets. A $\theta$-basis set $\mathbf{B} = \{B_1, B_2, \ldots, B_w\}$, where each $B_i$ is a set of items, has the property that any itemset with frequency higher than $\theta$ is a subset of some basis $B_i$. A good basis set is one where $w$ is small and the lengths of all $B_i$'s are also small. Given a good basis set $\mathbf{B}$, one can reconstruct the frequencies of all subsets of $B_i$'s with good accuracy. One can then select the most frequent itemsets from these. We also introduce techniques to construct good basis sets while satisfying differential privacy. Finally, we have conducted extensive experiments, and the results show that our approach greatly outperforms the existing approach.

We call our approach PrivBasis. It meets the challenge of high dimensionality by projecting the input dataset onto a small number of selected dimensions that one cares about. In fact, PrivBasis often uses several sets of dimensions for such projections, to avoid any one set containing too many dimensions. Each basis in $\mathbf{B}$ corresponds to one such set of dimensions for projection. Our techniques enable one to select which sets of dimensions are most helpful for the purpose of finding the $k$ most frequent itemsets.

The rest of this paper is organized as follows. The next section introduces the background knowledge about differential privacy and frequent itemset mining. In Section 3 we analyze the state of the art on private FIM and identify the challenges in this problem. Our approach is presented in Section 4. We report experimental results in Section 5. Section 6 reviews related work and Section 7 concludes our work.

## 2. PRELIMINARIES

### 2.1 Differential Privacy

Informally, differential privacy requires that the output of a data analysis mechanism be approximately the same, even if any single tuple in the input database is arbitrarily added or removed.

DEFINITION 1 ($\epsilon$-DIFFERENTIAL PRIVACY [16, 17]).
*A randomized mechanism $\mathcal{A}$ gives $\epsilon$-differential privacy if for any pair of neighboring datasets $D$ and $D'$, and any $S \in Range(\mathcal{A})$,*

$$\Pr[\mathcal{A}(D) = S] \le e^\epsilon \cdot \Pr[\mathcal{A}(D') = S].$$

In this paper we consider two datasets $D$ and $D'$ to be neighbors if and only if either $D = D' + t$ or $D' = D + t$, where $D + t$ denotes the dataset resulted from adding the tuple $t$ to the dataset $D$. We use $D \simeq D'$ to denote this. This protects the privacy of any single tuple, because adding or removing any single tuple results in $e^\epsilon$-multiplicative-bounded changes in the probability distribution of the output. If any adversary can make certain inference about a tuple based on the output, then the same inference is also likely to occur even if the tuple does not appear in the dataset.

Differential privacy is composable in the sense that combining multiple mechanisms that satisfy differential privacy for $\epsilon_1, \cdots, \epsilon_m$ results in a mechanism that satisfies $\epsilon$-differential privacy for $\epsilon = \sum_i \epsilon_i$. Because of this, we refer to $\epsilon$ as the privacy budget of a privacy-preserving data analysis task. When a task involves multiple steps, each step uses a portion of $\epsilon$ so that the sum of these portions is no more than $\epsilon$.

There are several approaches for designing mechanisms that satisfy $\epsilon$-differential privacy. In this paper we use two of them. The first approach computes a function $g$ on the dataset $D$ in a differentially privately way, by adding to $g(D)$ a random noise. The magnitude of the noise depends on $\mathsf{GS}_f$, the *global sensitivity* or the $L_1$ sensitivity of $g$. Such a mechanism $\mathcal{A}_g$ is given below:

$$\mathcal{A}_g(D) = g(D) + \mathsf{Lap}\left(\frac{\mathsf{GS}_g}{\epsilon}\right)$$
where $\mathsf{GS}_g = \max_{(D,D'):D \simeq D'} |g(D) - g(D')|,$
and $\Pr[\mathsf{Lap}(\beta) = x] = \frac{1}{2\beta} e^{-|x|/\beta}$

In the above, $\mathsf{Lap}(\beta)$ denotes a random variable sampled from the Laplace distribution with scale parameter $\beta$. This is generally referred to as the *Laplacian mechanism* for satisfying differential privacy.

The second approach computes a function $g$ on a dataset $D$ by sampling from the set of all possible answers in the range of $g$ according to an exponential distribution, with answers that are "more accurate" will be sampled with higher probability. This is generally referred to as the *exponential mechanism* [28]. This approach requires the specification of a *quality* function $q : \mathcal{D} \times \mathcal{R} \to \mathbb{R}$, where the real valued score $q(D, r)$ indicates how accurate it is to return $r$ when the input dataset is $D$. Higher scores indicate more accurate outputs which should be returned with higher probabilities. Given the quality function $q$, its global sensitivity $\mathsf{GS}_q$ is defined as:

$$\mathsf{GS}_q = \max_r \max_{(D,D'):D \simeq D'} |q(D,r) - q(D',r)|$$

The following method $\mathcal{M}$ satisfies $\epsilon$-differential privacy:

$$\Pr[\mathcal{M}(D) = r] \propto \exp\left(\frac{\epsilon}{2\,\mathsf{GS}_q} q(D,r)\right) \quad (1)$$

For example, if $q(D, r_1) - q(D, r_2) = 1$, then $r_1$ should be returns $y$ times more likely than $r_2$, with $y = \exp\left(\frac{\epsilon}{2\,\mathsf{GS}_q}\right)$. The larger the exponent $\frac{\epsilon}{2\,\mathsf{GS}_q}$ is, the more likely that $\mathcal{M}$ will return the higher quality result.

| Symbol | Description |
|---|---|
| $D$ | The transaction dataset |
| N | The number of transactions in $D$ |
| **I** | The set of items |
| **B** | The basis set |
| $f(X)$ | The frequency of itemset $X$ |
| $\lambda$ | The number of unique items in the set of top-$k$ itemsets |
| $f_k$ | The frequency of the $k$-th most frequent itemset |

**Table 1: The notations**

As pointed out in [28], in some cases the quality function satisfies the condition that when the input dataset is changed from $D$ to $D'$, the quality values of all outcomes change only in one direction, i.e.,

$$\forall_{D \simeq D'} \left[ \left( \exists_{r_1} q(D, r_1) < q(D', r_1) \right) \to \left( \forall_{r_2} q(D, r_2) \le q(D', r_2) \right) \right]$$

Then one can remove the factor of $1/2$ in the exponent of Equation (1) and return $r$ with probability proportional to $\exp\left(\frac{\epsilon}{\mathsf{GS}_q} q(D,r)\right)$. This improves the accuracy of the result.

## 2.2 Frequent Itemset Mining

Frequent itemset mining (FIM) is a well studied problem in data mining. It aims at discovering the itemsets that frequently appear in a transactional dataset. More formally, let **I** be a set of items and let $D = [t_1, t_2, \ldots, t_N]$ be a transaction dataset where $t_i \subseteq \mathbf{I}$, and $N$ be the number of transactions in $D$. The *frequency* of an itemset $X \subseteq \mathbf{I}$, denoted by $f(X)$, is the fraction of transactions in $D$ that include $X$ as a subset. Thus $0 \le f(X) \le 1$ for any $X$. Given a frequency threshold $\theta$ such that $0 \le \theta \le 1$, we say that an itemset $X$ is $\theta$-frequent when $f(X) \ge \theta$.

The FIM problem can be defined as either taking the minimal frequency $\theta$ as input and returning all $\theta$-frequent itemsets together with their frequencies, or as taking an integer $k$ as input, and returning the top $k$ most frequent itemsets, together with their frequencies. One can easily convert one version to the other.

Several algorithms have been proposed for finding frequent itemsets. The two most prominent ones are the Apriori algorithm [5], and the FP-Growth algorithm [22]. The Apriori algorithm exploits the observation that if an itemset $X$ is frequent, then all its subsets must also be frequent. The algorithm works by generating itemsets of length $n$ from itemsets of length $n - 1$, eliminating candidates that have an infrequent pattern. The FP-Growth algorithm skips the candidate itemset generation process by using a compact tree structure to store itemset frequency information.

## 3. THE EXISTING APPROACH

This paper is inspired by Bhaskar et al.'s paper [8] in KDD'10, which proposed an approach for releasing the top $k$ itemsets of a predefined length $m$. That is, among all itemsets of length **exactly** $m$, one chooses the $k$ most frequent ones, and releases their frequencies. This method can be easily extended to the case of releasing top $k$ itemsets of length **at most** $m$, instead of exactly $m$. This can then be used to return top $k$ most frequent itemsets by choosing an appropriate $m$. To the best of our knowledge, this is the only existing approach that publishes frequent itemsets while satisfying differential privacy. A key concept introduced in this approach is the notion of Truncated Frequencies, we thus call this approach the TF method.

The TF method has two steps. In the first step, one selects which $k$ itemsets to release. In the second step, one releases the frequencies of these itemsets after adding noises to them. The privacy budget $\epsilon$ is evenly divided between these two steps. The second step is straightforward. Given $k$ itemsets, releasing their frequencies have sensitivity $\frac{k}{N}$, as adding or removing one transaction can affect the frequency of each itemset by at most $\frac{1}{N}$. Thus adding noise according to $\mathsf{Lap}\left(\frac{2k}{N\epsilon}\right)$ to the frequency of each of the $k$ itemsets satisfies $(\epsilon/2)$-differential privacy.

The main challenge lies in the first step, namely, selecting the $k$ most frequent itemsets. The TF method selects these from the set of all itemsets of length at most $m$; we use $U$ to denote this set of candidate itemsets. The number of these candidate itemsets is

$$|U| = \sum_{i:1\leq i\leq m} \binom{|\mathbf{I}|}{i} \approx |\mathbf{I}|^m, \quad (2)$$

which is exponential in $m$.

Because $|U|$ is large, enumerating through all elements in $U$ can be computationally expensive. The key novelty in the TF approach is to use the notion of truncated frequencies to avoid explicitly enumerating through all itemsets in $U$. The truncated frequency of $X \in U$ is defined as $\hat{f}(X) = \max(f(X), f_k - \gamma)$, where $f_k$ is the frequency of the $k$'th most frequent itemset in $U$, and $\gamma$ is a parameter computed using Equation (3) below.

The intuition is that for an itemset with frequency below $f_k - \gamma$, one does not need to explicitly consider the itemset; instead, it suffices to use $f_k - \gamma$ as the upperbound of the itemset's frequency. The parameter $\gamma$ must be selected to ensure that an itemset of frequency less than $f_k - \gamma$ is selected only with low probability; it is computed as follows:

$$\gamma = \frac{4k}{\epsilon N}\left(\ln\frac{k}{\rho} + \ln|U|\right) \quad (3)$$

The value $\rho$ bounds the error probability and should be between 0 and 1.

Two methods were proposed to select the $k$ most frequent itemsets from $U$ using the truncated frequencies. The first is to add $\mathsf{Lap}\left(\frac{4k}{\epsilon N}\right)$ to the truncated frequencies of all itemsets in $U$, and then select the $k$ with highest noisy frequencies. The second method is to use repeated applications of the Exponential Mechanism. One samples $k$ times without replacement, such that the probability of selecting an itemset, $X$, is proportional to $\exp\left(\frac{\epsilon N}{4k}\hat{f}(X)\right)$. It is shown that both methods satisfy $(\epsilon/2)$-differential privacy. For both methods, the algorithm explicitly considers only the itemsets with frequencies $> f_k - \gamma$, and estimates the probability that it should select an itemset whose frequency is $\leq f_k - \gamma$ (i.e., with truncated frequency $= f_k - \gamma$), and then randomly samples such an itemset.

Furthermore, it is proven that the output of either method provides the following utility guarantee: With probability $1-\rho$, every itemset with true frequency at least $f_k + \gamma$ are selected, and every selected itemset has true frequency at least $f_k - \gamma$.

## 3.1 Analysis of the TF Method

The TF method works well when $k$ is small. However, it scales poorly with larger $k$. To see why this is the case, recall that the TF method enumerates only itemsets with frequencies above $f_k - \gamma$ to prune the search space. When $f_k - \gamma \leq 0$, i.e, when $\gamma \geq f_k$, this technique results in no pruning at all, and the algorithm degenerates into explicitly enumerating through all elements of $U$. Furthermore, the proven utility guarantee (e.g., every selected itemset has frequency at least $f_k - \gamma$) is meaningless when $f_k - \gamma \leq 0$.

Unfortunately, as Table 2(a) shows, in many datasets with large $k$ ($k \geq 100$, or $k \geq 200$), we have $\gamma$ larger than, or very close to $f_k$. To see why, observe that

$$\gamma = \frac{4k}{\epsilon N}\left(\ln\frac{k}{\rho} + \ln|U|\right) > \frac{4k}{\epsilon N}\left(\ln|U|\right) \approx \frac{4km\ln|\mathbf{I}|}{\epsilon N}$$

That is, $\gamma$ grows linearly in $km$. When $k$ is large, the top $k$ itemsets likely include many itemsets of sizes 3, 4, or higher. If one chooses a small $m$, then one misses all frequent itemsets that of size greater than $m$. If one chooses a larger $m$, e.g., 3 or higher, then the $\gamma$ value is too large, rendering the mechanism unfeasible.

We observe that a deeper reason why the TF method does not scale is that when one needs to select the top $k$ itemsets from a large set $U$ of candidates, the large size of $U$ causes two difficulties. The first is regarding the *running time*, i.e., a large $|U|$ makes enumerating through all elements in $U$ unfeasible. The second difficulty is about *accuracy*, i.e., a large $|U|$ makes the selection of top $k$ candidates from $U$ inaccurate. Even if every *single* low-frequency itemset in $U$ is chosen with only a small probability, the sheer number of such low-frequency itemsets means that the $k$ selected itemsets likely include many infrequent ones. The TF technique tries to address the running time challenge by pruning the search space, but it *does not* address the accuracy challenge. This addresses only one symptom caused by a larger candidate set, but not the root cause. In the end, even the goal of improving running time cannot be achieved when $|U|$ is large, because the accuracy requirement forces a large $\gamma$.

## 4. THE PRIVBASIS METHOD

In this section, we introduce the PrivBasis method for publishing the top $k$ frequent itemsets. If one desires to publish all itemsets above a given threshold $\theta$, one can compute the value $k$ such that the $k$'th most frequent itemset has frequency $\geq \theta$ and the $k+1$'th itemset has frequency $< \theta$, and then uses PrivBasis to find the top $k$ frequent itemsets.

### 4.1 Overview of PrivBasis

We observe that the key challenge of dealing with transaction datasets is their high dimensionality. The PrivBasis approach can be viewed as meeting the challenge by projecting the input dataset $D$ onto lower dimensions. For example, let $B$ be the set of $\ell$ most frequent items in $D$; projecting $D$ to items in $B$ means removing from every transaction all the items that are not in $B$. The $\ell$ items in $B$ can be viewed as $\ell$ binary attributes that partition the dataset into $2^\ell$ bins. Using the standard Laplacian mechanism, one can obtain the noisy frequency of each bin, through which one can reconstruct the frequencies of all subsets of $B$. For this method to work, the value $\ell$ cannot be much larger than

| dataset | $N$ | $|\mathbf{I}|$ | avg $|t|$ | $k$ | $\lambda$ | $\lambda_2$ | $\lambda_3$ |
|---|---|---|---|---|---|---|---|
| retail | 88162 | 16470 | 11.3 | 100 | 38 | 37 | 21 |
| mushroom | 8124 | 119 | 24 | 100 | 11 | 30 | 36 |
| pumsb-star | 49046 | 2088 | 50 | 200 | 17 | 31 | 50 |
| kosarak | 990002 | 41270 | 8.1 | 200 | 39 | 84 | 58 |
| AOL | 647377 | 2290685 | 34 | 200 | 171 | 29 | 0 |

(a) Dataset parameters: avg $|t|$ is the average transaction length, $\lambda$ is the number of unique items in the top $k$ itemsets, and $\lambda_2$ ($\lambda_3$, resp.) is the number of pairs (size-3 itemsets, resp.) in the top $k$ itemsets. Note that we choose $\rho = 0.9$, which requires utility guarantee only with the low probability of $1 - \rho = 0.1$.

| dataset | $k$ | $\mathbf{f_k \cdot N}$ | $m$ | $|U| \approx \binom{|I|}{m}$ | $\gamma \cdot \mathbf{N}$ |
|---|---|---|---|---|---|
| retail | 100 | **1192** | 1 | $16{,}470$ | **5768** |
| mushroom | 100 | **4464** | 2 | $7{,}104$ | **5433** |
| pumsb-star | 200 | **28613** | 3 | $1.5 \times 10^9$ | **21235** |
| kosarak | 200 | **14142** | 2 | $8.5 \times 10^8$ | **20733** |
| AOL | 200 | **12450** | 1 | $2.3 \times 10^6$ | **16038** |

(b) Effectiveness of the TF approach due to Bhaskar et al., when applied to selecting top $k$ itemsets that are of length at most $m$. When the column $\gamma \cdot N$ is larger than $f_k \cdot N$, the truncated frequency approach is completely ineffective.

Table 2: Effectiveness of the TF approach, and dataset parameters

a dozen. However, for many datasets, when recovering the top $k$ itemsets for $k = 100$, $k = 200$, or even larger $k$, one needs to go beyond the first dozen or so most frequent items. Thus one needs to choose more than one sets of items (i.e., dimensions) for projections. How to select these in a differentially private fashion, and how to best utilize information obtained from them are the main challenges that need to be solved by the PrivBasis method.

Formalizing the above intuitions, we introduce the concept of $\theta$-basis sets of a transaction dataset.

DEFINITION 2 ($\theta$-BASIS SET). *Given a transaction dataset $D$ over items in $\mathbf{I}$, and a threshold $\theta$, we say that $\mathbf{B} = \{B_1, B_2, \ldots, B_w\}$, where $B_i \subseteq \mathbf{I}$ for $1 \leq i \leq w$, is a $\theta$-basis set for $D$, if and only if for any $\theta$-frequent itemset $X \subseteq \mathbf{I}$, there exists $B_i \in \mathbf{B}$, such that $X \subseteq B_i$. We say that $B_i$ covers $X$.*

*We call $w$ the width of the basis set, $\ell = \max_{1 \leq i \leq w} |B_i|$ the length of the basis set, and each $B_i$ a basis.*

Given a dataset $D$ and a $\theta$-basis set $\mathbf{B}$ for it, we can privately reconstruct with reasonable accuracy the frequencies of all itemsets in the the following candidate set $\mathcal{C}(\mathbf{B})$.

DEFINITION 3 (CANDIDATE SET). *The candidate set given a $\theta$-basis $\mathbf{B} = \{B_1, B_2, \ldots, B_w\}$ is defined as*

$$\mathcal{C}(\mathbf{B}) = \bigcup_{i=1}^{w} \{X | X \subseteq B_i\}$$

That is, the candidate set $\mathcal{C}(\mathbf{B})$ is the set of all itemsets that are covered by some basis $B_i$ in $\mathbf{B}$. When $\mathbf{B}$ is a $\theta$-basis set, all $\theta$-frequent itemsets are in $\mathcal{C}(\mathbf{B})$.

At a high level, the PrivBasis method consists of the following steps.

1. Obtain $\lambda$, the number of unique items that are involved in the $k$ most frequent itemsets.

2. Obtain $F$, the $\lambda$ most frequent items among $\mathbf{I}$. The desired goal (which can be approximately achieved) is that $F$ includes exactly the items that appear in the top $k$ itemsets.

3. Obtain $P$, a set of the most frequent pairs of items among $F$. The desired goal is that $P$ includes exactly the pairs of items that appear in the top $k$ itemsets.

4. Construct $\mathbf{B}$, using $F$ and $P$. The desired goal is that $\mathbf{B}$ is a $f_k$-basis set with small width and length.

5. Obtain noisy frequencies of itemsets in $\mathcal{C}(\mathbf{B})$; one can then select the top $k$ itemsets from $\mathcal{C}(\mathbf{B})$.

In the rest of this section, we present details of these steps. We do this in a reverse order, first presenting Step 5 in Section 4.2, then Step 4 in Section 4.3, and finally the complete algorithm, including details of Steps 1 to 3 in Section 4.4.

## 4.2 Generating Noisy Counts for $\mathcal{C}(\mathbf{B})$

Algorithm 1 gives the BASISFREQ algorithm for computing the noisy counts of all itemsets in $\mathcal{C}(\mathbf{B})$. In the algorithm we compute noisy counts, which can be translated into frequencies easily. The key ideas of the algorithm are as follows. Each basis $B_i$ divides all possible transactions into $2^{|B_i|}$ mutually disjoint bins, one corresponding to each subset of $B_i$. For each $X \subseteq B_i$, the bin corresponding to $X$ consists of all transactions that contain all items in $X$, but no item in $B_i \setminus X$.

Given a basis set $\mathbf{B}$, adding noise $\mathsf{Lap}(w/\epsilon)$ to each bin count and outputting these noisy counts satisfy $\epsilon$-differential privacy. For each basis $B_i$, adding or removing a single transaction can affect the count of exactly one bin by exactly 1. Hence the sensitivity of publishing all bin counts for one basis is 1; and the sensitivity for publishing counts for all bases is $w$. In Algorithm 1, lines 2 to 11 compute these noisy bin frequencies.

From these frequencies, one can then recover the counts of all itemsets in $\mathcal{C}$. For example, a basis $\{a, b, c\}$ divides all transactions into 8 bins: $\{\neg a, \neg b, \neg c\}$ (not containing any of $a, b, c$), $\{a, \neg b, \neg c\}, \cdots, \{a, b, c\}$. The count of the itemset $\{a, b\}$ can then be obtained by summing up the counts for the two bins $\{a, b, \neg c\}$ and $\{a, b, c\}$. Lines 12 to 26 in Algorithm 1 compute the noisy counts for itemsets in $\mathcal{C}$.

THEOREM 1. *Algorithm 1 is $\epsilon$-differentially private.*

PROOF. The only part in Algorithm 1 that depends on the dataset is computing the noisy bin counts $b[i][X]$. As discussed above, publishing all bin counts has sensitivity $w/N$. Line 4 adds Laplacian noise to satisfy $\epsilon$-differential privacy taking this sensitivity into consideration. Starting from line 12, the algorithm only performs post-processing, and does not access $D$ again. □

**Running Time.** We now analyze the running time of Algorithm 1. The algorithm has four parts. The first part

**Algorithm 1 BasisFreq**: Privately Releasing Frequent Itemsets using Basis Sets

**Input:** Transactional dataset $D$, $\mathbf{B} = \{B_1, \cdots, B_w\}$, $k$, differential privacy budget $\epsilon$.
**Output:** Top $k$ frequent itemsets in $\mathcal{C}$ and their frequencies.

1: **function** BASISFREQ($D, \mathbf{B} = \{B_1, \cdots, B_w\}, k, \epsilon$)
2:    **for** $i = 1 \to w$ **do**
3:       **for** $j = 0 \to 2^{|B_i|} - 1$ **do**
4:          $b[i][j] \leftarrow \mathsf{Lap}\left(\frac{w}{\epsilon}\right)$
5:       **end for**
6:    **end for**
7:    **for all** $t \in D$ **do**
8:       **for** $i = 1 \to w$ **do**
9:          $b[i][t \cap B_i] \leftarrow b[i][t \cap B_i] + 1$
10:       **end for**
11:    **end for**
12:    $C \leftarrow \emptyset$
13:    **for** $i = 1 \to w$ **do**
14:       **for all** $X \subseteq B_i$ **do**
15:          $nc \leftarrow \sum_{Y \subseteq B_i | X \subseteq Y} b[i][Y]$
16:          $nv \leftarrow 2^{|B_i| - |X|}$
17:          **if** $C(X)$ is undefined **then**
18:             $C(X).nc \leftarrow nc$
19:             $C(X).v \leftarrow nv$
20:          **else**
21:             $v \leftarrow C(X).v$
22:             $C(X).nc \leftarrow \frac{nv}{v+nv} C(X).nc + \frac{v}{v+nv} nc$
23:             $C(X).v \leftarrow \frac{v \cdot nv}{v+nv}$
24:          **end if**
25:       **end for**
26:    **end for**
27:    $R \leftarrow$ the $k$ elements in $X$'s with highest $C(X).nc$
28:    **return** $R$
29: **end function**

**Comments**

- The array element $b[i][X]$ stores the noisy count of the bin corresponding to itemset $X$ and basis $B_i$. We point out that a subset of $B_i$ is easily converted to binary number of $|B_i|$ bits, which is an integer index in $0..2^{|B_i|} - 1$.
- The rationale behind lines 21-23 are explained in Section 4.2 in the "Noisy Frequency Accuracy Analysis" paragraph.

(lines 2 to 6) initializes the array $b$, which takes $O(w2^\ell)$ time. The second part (lines 7 to 11) scans the dataset $D$ and matches each transaction with each basis, and takes time $O(w|D|)$, where $|D|$ is the sum of the lengths of all transactions in $D$. The third part (lines 12 to 26) computes the noisy counts of itemsets in the candidate set $\mathcal{C}$. This part's runtime is dominated by line 15, which for each $X \subseteq B_i$ requires $O(2^{|B_i|-|X|})$ operations. The total number of operations involved for each basis $B_i$ is

$$\sum_{j=1}^{|B_i|} \binom{|B_i|}{j} 2^{|B_i|-j} = 3^{|B_i|} - 2^{|B_i|}$$

Thus the third part takes time $O(w3^\ell)$. The last part (line 27) sorts the noisy counts of itemset in $\mathcal{C}$ to select top $k$ and takes time $O(w\ell(\log w)2^\ell)$. Thus Algorithm 1 has time complexity $O(w|D| + w3^\ell)$. For large dataset, we will have $\ell < \log_3 |D|$, and the running time is dominated by $O(w|D|)$. This analysis shows that $w$ is a linear factor on the running time, while $\ell$ has an exponential effect. In our experiments we limit $\ell$ to be at most 12, and often 10 or smaller.

**Accuracy Analysis.** We now analyze the accuracy of the noisy frequencies obtained via Algorithm 1. Let $nf_i(X)$ denote the noisy frequency of an itemset $X$ from a basis $B_i$. We use the Error Variance as the measure of accuracy. That is, we consider

$$\mathsf{EV}[nf_i(X)] = \mathsf{Var}(|nf_i(X) - f(X)|)$$

When computing $nf_i(X)$, one sums up $2^{|B_i|-|X|}$ noisy counts, each with noise independently generated according to $\mathsf{Lap}\left(\frac{w}{\epsilon N}\right)$, which has variance $\frac{2w^2}{\epsilon^2 N^2}$. We thus have:

$$\mathsf{EV}[nf_i(X)] = 2^{|B_i|-|X|+1} \frac{w^2}{\epsilon^2 N^2} \qquad (4)$$

When two or more bases overlap, some itemsets may be subsets of more than one bases, and one thus obtains multiple noisy counts of such an itemset, one from each basis including the itemset. In this case, these counts can be combined to obtain a more accurate count. Given two noisy counts of the itemset $X$, $nf_1$ with error variances $v_1$ and $nf_2$ with error variance $v_2$, the optimal way to combine them is to use $\frac{v_2}{v_1+v_2} nf_1 + \frac{v_1}{v_1+v_2} nf_2$, resulting in error variance $\frac{v_1 v_2}{v_1+v_2}$. This weighted averaging is done in lines 22 and 23.

From Equation (4), it is easy to see that the worst-case error variance among all $X$ and all $B_i$ is $2^\ell \frac{w^2}{\epsilon^2 N^2}$. To minimize such worst-case error variance, one wants to minimize $w^2 2^\ell$.

Alternatively, one may want to minimize average-case error variance. Given a basis set $\mathbf{B}$, and a set $Q$ of itemsets, we can compute the average-case error variance for using $\mathbf{B}$ to obtain noisy frequencies for itemsets in $Q$ as follows. For each itemset $X \in Q$, if it is covered by a single basis $B_i$, then the error variance can be computed using Equation 4. If $X$ is covered by more than one bases, one can also computed the error variance of the weighted average method. One can then take an average of the computed error variance for all itemsets in $Q$.

We now consider a special case where one wants to obtain noisy frequencies for a set $Q$ of $k$ individual items, and show what basis set minimizes both the worst-case and the average-case error variance. One extreme is to use one basis for each item in $Q$. As a result, one adds Laplacian noise to the $k$ counts, and has sensitivity $k$. The noise has distribution $\mathsf{Lap}\left(\frac{k}{\epsilon}\right)$, resulting in error variance of $k^2 V$, where $V = \frac{2}{\epsilon^2}$. Now assume that the $k$ items are divided into bases of size $\ell$, then we have $w = k/\ell$ bases. The noise variance for the frequency of each item is thus

$$w^2 2^{\ell-1} V = \frac{2^{\ell-1}}{\ell^2} k^2 V.$$

Note that $\frac{2^{\ell-1}}{\ell^2}$ is minimized at $\ell = 3$, where it equals $4/9$. Thus one obtains more than half reduction in the error variance when compared with the direct method.

## 4.3 Constructing Basis Sets

We now discuss how to construct a good $\theta$-basis set for $\theta = f_k$, the frequency of the $k$-th most frequent itemset.

Given a transaction dataset $D$ and a threshold $\theta$, many $\theta$-basis sets exist. As discussed above, we desire $\theta$-basis sets that have both small width and small length. We now investigate some properties of $\theta$-basis sets, which help us construct $\theta$-basis sets.

PROPOSITION 2. *The set* $\mathbf{B_1} = \{\{x_1, \cdots, x_\lambda\}\}$, *where $x_1, \cdots, x_\lambda$ are all the items that are $\theta$-frequent, is a $\theta$-basis set of $D$ of width 1 and length $\lambda$.*

PROOF. For any $\theta$-frequent itemset $X$, all items in $X$ must be $\theta$-frequent, and therefore $X \subseteq B_1$. □

We use $\lambda$ to denote the number of unique items in the top $k$ itemsets. When $\lambda$ is small, then this basis-set would work fine. However, when $\lambda$ is large, we need other methods. Below we explore additional properties of $\theta$-basis sets.

PROPOSITION 3. *The set* $\mathbf{B_m} = \{B_1, B_2, \cdots, B_w\}$, *where $B_1, B_2, \cdots, B_w$ are all the maximal $\theta$-frequent itemsets, is a $\theta$-basis set. Furthermore, any $\theta$-basis set must have length at least as large as the length of $\mathbf{B_m}$.*

PROOF. Recall that a maximal frequent itemset is a $\theta$-frequent itemset such that any superset of it is not $\theta$-frequent. The length of this $\theta$-basis set equals the size of the largest maximal $\theta$-frequent itemset. This $\theta$-basis set has the smallest length among all $\theta$-basis sets, because every $\theta$-basis set must include a basis that is a superset of the largest maximal $\theta$-frequent itemset. □

PROPOSITION 4. *Given a $\theta$-basis set $\mathbf{B} = \{B_1, B_2, \ldots, B_w\}$, merging any $B_i$ and $B_j$ (i.e., replace $B_i$ and $B_j$ with $B_i \cup B_j$ in $\mathbf{B}$) results in another $\theta$-basis set of width $w - 1$.*

Propositions 3 and 4 together give one way to construct good basis sets. Given the maximal frequent itemsets, one can merge them as needed. The challenge is that it is unclear how to publish the set of all maximal frequent itemsets while satisfying differential privacy. Below we show a way to privately over-approximate the maximal frequent itemsets.

DEFINITION 4. *Let $F$ be the set of $\theta$-frequent items, and $P$ be the set of all $\theta$-frequent pairs of items. Observe that $P$ involves only items in $F$. We define the $\theta$-**frequent pairs graph** to be the graph where each node corresponding to an item in $F$ and each edge corresponds to a frequent pair in $P$.*

We are interested in the maximal *maximal cliques* in the $\theta$-frequent graph. A maximal clique, sometimes called inclusion-maximal, is a clique that is not included in a larger clique. The classic algorithm for finding all maximum cliques is the Bron-Kerbosch Algorithm [12], which is widely used in application areas of graph algorithms.

PROPOSITION 5. *Given $D$, the set of all maximal cliques in $D$'s $\theta$-frequent pairs graph form a $\theta$-basis-set of $D$.*

PROOF. For any $\theta$-frequent item $x$, $x \in F$ belongs to some maximal clique. For any $\theta$-frequent itemset $X$ of size $\geq 2$, by the apriori principle all items in $X$ and all pairs of items in $X$ must be $\theta$-frequent, thus $X$ corresponds to a clique in the $\theta$-frequent pairs graph, which must be included in some maximal clique. □

The above proof also shows that each $\theta$-frequent itemset must be a subset of some maximal clique of the $\theta$-frequent pair graph; however, a maximal clique may not be a $\theta$-frequent itemset. For example, it may be the case that pairs $\{1, 2\}, \{2, 3\}, \{3, 4\}$ are all $\theta$-frequent, but the itemset $\{1, 2, 3\}$ is not $\theta$-frequent. Thus, the maximal cliques over-approximate the maximal frequent itemsets.

We use Propositions 5 and 4 to construct a basis set. Algorithm 2 gives the algorithm for constructing a basis set that covers all maximal cliques in the graph constructed from $F$ and $P$, while attempting to minimize the average-case error variance (EV) for pairs in $P$ and items in $F \setminus P$, which we use to denote items that appear in $F$, but not in $P$.

The algorithm starts with a basis set that has two parts: $\mathbf{B_1}$ includes the maximal cliques of size at least 2; $\mathbf{B_2}$ includes items in $F \setminus P$ grouped into itemsets of size 3 each, with possibly 1 or 2 items left. The algorithm then greedily merges bases in $\mathbf{B_1}$, to reduce the EV. After this step, the algorithm tries to remove some basis in $\mathbf{B_2}$ can distribute the items in them elsewhere, if doing so reduces the EV.

---

**Algorithm 2** ConstructBasisSet: Construct a Basis Set Using Frequent Items and Pairs

**Input:** $F$, frequent items, and $P$, frequent pairs.
**Output:** $\mathbf{B}$, a basis set covering all maximal cliques in the graph $(F, P)$.

1: **function** CONSTRUCTBASISSET($F, P$)
2:    $\mathbf{B_1} \leftarrow$ all maximal cliques of size at least 2 in the graph given by $P$
3:    $\mathbf{B_2} \leftarrow$ items in $F$ but not in $P$, divided into the smallest number of itemsets such that each contains at most 3 items
4:    Repeatedly find $B_i, B_j \in \mathbf{B_1}$ such that merging $B_i$ and $B_j$ results in the largest reduction of average-case error variance (EV) when using $\mathbf{B} = \mathbf{B_1} \cup \mathbf{B_2}$ to obtain frequencies of itemsets in $F$ and $P$; and update $\mathbf{B_1}$ by merging $B_i, B_j$; stop when no merging reduces EV
5:    Repeatedly find $B_i \in \mathbf{B_2}$ such that removing $B_i$ and moving items in $B_i$ to bases in $\mathbf{B_1} \cup \mathbf{B_2}$ with smallest sizes results in the largest EV-reduction; update $\mathbf{B}$ when $B_i$ is found; stop when no such $B_i$ can be found
6:    **return** $\mathbf{B} = \mathbf{B_1} \cup \mathbf{B_2}$
7: **end function**

---

## 4.4 Putting Things Together for PrivBasis

Now we are able to put all the pieces together for the PrivBasis method. The algorithm is given in Algorithm 3.

Recall that the algorithm has five steps, as given in Section 4.1: (1) Get lambda; (2) Get frequent items; (3) Get frequent pairs; (4) Construct the basis set; (5) Get noisy counts.

**Privacy Budget Allocation.** The privacy budget $\epsilon$ must be divided among the steps $1, 2, 3, 5$. Step 4 does not access the dataset $D$, and only processes the outputs of earlier steps. We divide the privacy budget into three portions: $\alpha_1\epsilon$ is used for Step 1 (obtaining $\lambda$), $\alpha_2\epsilon$ is used for Steps 2 and 3 combined, $\alpha_3\epsilon$ is used for Step 5. In our experiments, we chose $\alpha_1 = 0.1$, $\alpha_2 = 0.4$, and $\alpha_3 = 0.5$ for all datasets. These choices were not tuned, and may not be

1345

**Algorithm 3 PrivBasis**: Privately Releasing Frequent Itemsets

**Input:** Transactional dataset $D$, items $\mathbf{I}$, $k$, differential privacy budget $\epsilon$.

**Algorithmic Parameters:** $\alpha_1 + \alpha_2 + \alpha_3 = 1$ decides what proportions of the privacy budget are allocated to the different steps. We use $\alpha_1 = 0.1, \alpha_2 = 0.4, \alpha_3 = 0.5$. Parameter $\eta$, which we set at either 1.1 or 1.2, is the safety margin parameter.

```
 1: function PRIVBASISMAIN(D, I, k, ε)
 2:     λ ← GETLAMBDA(D, I, k, α₁ε)
 3:     if λ ≤ 12 then
 4:         F ← GETFREQITEMS(D, I, λ, α₂ε)
 5:         return BASISFREQ(D, {F}, k, (1 − α₁ − α₂)ε)
 6:     else
 7:         λ₂ = η · k − λ
 8:         λ₂ ← λ₂/√max(1, λ₂/λ)
 9:         β₁ ← α₂ · λ/(λ + λ₂)
10:         β₂ ← α₂ − β₁
11:         F ← GETFREQELEMENTS(D, I, λ, β₁ε)
12:         U ← all pairs of items in F
13:         P ← GETFREQELEMENTS(D, U, λ₂, β₂ε)
14:         B = CONSTRBASIS(F, P)
15:         return BASISFREQ(D, B, k, (1 − α₁ − α₂)ε)
16:     end if
17: end function

18: function GETLAMBDA(D, I, k, ε)
19:     N ← number of transactions in D
20:     k₁ ← ⌈k · η⌉
21:     θ ← frequency of k₁'th itemset
22:     for i = 1 → |I| do
23:         f ← frequency of i'th item
24:         p[i] ← e^((1−|f−θ|)·N·ε/2)
25:     end for
26:     λ ← sample i ∈ [1..|I|] according to p[i]
27:     return λ
28: end function

29: function GETFREQELEMENTS(D, U, λ, ε)
30:     N ← number of transactions in D
31:     for i = 1 → |U| do
32:         f ← frequency of U[i]
33:         p[i] ← e^(f·ε/λ)
34:     end for
35:     for i = 1 → λ do
36:         X[i] ← sample from U[i] according to p[i]
37:         remove X[i] from U
38:     end for
39:     return X
40: end function
```

optimal; it appears that the optimal allocation depends on characteristics of the dataset $D$ and the value $k$.

Step 3 is needed only when $\lambda$, the number of unique items that appear in the top $k$ frequent itemsets, is $> 12$. Recall that when $\lambda \leq 12$, we construct $\mathbf{B}$ to consist of a single basis that includes the $\lambda$ most frequent items, and do not need step (3) to obtain frequent pairs. When $\lambda$ is small, we let Step 2 use the whole of $\alpha_2\epsilon$. When Step 3 is needed, the privacy budget $\alpha_2\epsilon$ must be allocated between Steps 2 and 3. This allocation is done according to how many frequent items and pairs we want to get. To obtain $\lambda$ most frequent items, and $\lambda_2$ most frequent pairs, Step 2 gets $\lambda/(\lambda + \lambda_2)$ portion of $\alpha_2\epsilon$, and Step 3 gets the rest.

**Step 1: Get $\lambda$.** Step 1 is done using the GETLAMBDA function in Algorithm 3. Intuitively, one can use the exponential method to sample $j$ from $\{1, 2, \cdots, k\}$ with the following quality function.

$$q(D, i) = (1 - |f_k - \mathit{fitem}_j|)N$$

where $\mathit{fitem}_j$ is the frequency of the $j$'th most frequent item. That is, we want to choose $j$ such that the $j$'th most frequent item has frequency closest to that of the $k$'th most frequent itemset.

The sensitivity of the above quality function is 1, because adding or removing a transaction can affect $f_k$ by at most $1/N$ and $\mathit{fitem}_j$ by at most $1/N$. Furthermore, $f_k$ and $\mathit{fitem}_j$ cannot change in different directions (i.e., one increases while the other decreases).

In Algorithm 3, rather than using $f_k$ in the above quality function, we use $f_{k_1}$, where $k_1 = k \cdot \eta$, and $\eta$ is a safety margin parameter that we set at either 1.1 or 1.2, depending on $k$. The reason for doing this is to avoid the error in which the obtained $\lambda$ is too small, because then we may miss a significant number of top $k$ itemsets with basis set constructed with top $\lambda$ items. When the obtained $\lambda$ is slightly larger than the correct value, this will just cause the privacy budget to be divided somewhat thinner, an effect we can tolerate better.

**Steps 2 and 3: Get frequent items and pairs.** Both Steps 2 and 3 use the GETFREQELEMENTS function, which privately selects a number of itemsets with highest frequencies from a set $U$. It uses repeated sampling without replacement, where each sampling step uses the exponential method with the frequency of each itemset as its quality.

In Step 2, we are selecting from all items in $I$, thus the candidate set size is $|I|$, the resulting set is $F$. In Step 3, we only need to select pairs of items in $F$; thus, the set $U$ from which we are selecting has only $\binom{\lambda}{2}$ elements, which is quite small.

When determining $\lambda_2$, the number of frequent pairs in the top $k$ itemsets, the naive method is to set $\lambda_2 = \eta \cdot k - \lambda$. This, however, is not ideal. In Table 2(b), we see that for the pumsb-star dataset, the top 100 itemsets include 17 items and 31 pairs. We desire a $\lambda_2$ value to be larger than 31, but not too large. Setting $\lambda_2 = \eta \cdot k - \lambda$ results a value close to 100 for $\eta = 1.2$. Obtaining top 100 pairs and constructing basis to cover them is inaccurate, both because each pair must be selected with less privacy budget, and because having to cover 100 pairs results in larger basis. While the best value of $\lambda_2$ depends on the dataset, we use the following heuristic formula.

$$\lambda_2 \leftarrow \frac{\lambda_2'}{\sqrt{\max(1, \lambda_2'/\lambda)}}, \text{ where } \lambda_2' = \eta \cdot k - \lambda$$

The intuition is that when the ratio of $\lambda_2'/\lambda$ is large, then we expect that a significant proportion of the top $\lambda$ items to be non-pairs, so we divide $\lambda_2'$ by the square root of the ratio. For the pumsb-star dataset, when the noisy $\lambda = 20$, the $\lambda_2$ value computed as above equals 44.

As all data-dependent step in Algorithm 3 satisfies differential privacy, we have the following theorem.

Theorem 6. *Algorithm 3 is $\epsilon$-differentially private.*

## 5. EXPERIMENTS

In this section, we demonstrate the efficacy of our approach through extensive experimental evaluation on a number of real datasets. We begin by describing the datasets we use as well as the utility measures we employ. We then present our experimental results.

**Datasets** To facilitate experimental evaluation, we run our algorithm on the following 5 datasets. Table 2(b) on page provides a description of the number of transactions in each datasets, the number of distinct items, as well as the average transaction length.

- *Retail Dataset* [2]. This is a retail market basket data from an anonymous Belgian retail store. Each transaction in the dataset is a set of items in one receipt and there are 88,162 receipts in total.

- *Mushroom Dataset* [2]. In this dataset, each record describes the physical attributes such as color of a single mushroom.

- *AOL Search Log Dataset* [1]. Each line of this dataset contains the randomly assigned userID, the search query string, the time stamp and the clicked URL. We preprocess the logs by removing the stop words and performing word stemming. By treating each query keyword as an item, we transform the preprocessed dataset into a transaction dataset by grouping the search query keywords of the same userID into one transaction.

- *Pumsb_star Dataset* [2]. The Pumsb dataset is census data from PUMS (Public Use Microdata Sample). Pumsb_star represents a subset of this dataset suitable for data mining purposes.

- *Kosarak Dataset* [2]. Kosarak contains the click-stream data of a Hungarian online news website. Each transaction is a click stream from a user.

**Utility Measures** We evaluate the utility by employing the following standard metrics.

- *False negative rate*: This measures the fraction of actual frequent itemsets which do not appear in the published result

$$FNR = \frac{\textsf{FalseNegatives}}{k}.$$

We point out that this is the same as the False Positive Rate, the fraction of identified top $k$ itemsets that are not in the actual top $k$.

- *Relative error of published itemset counts*: This measures the error with respect to the actual itemset frequency in the dataset. This is calculated over all published frequent itemsets.

$$RE = \textsf{median}_X \frac{|nf(X) - f(X)|}{f(X)}.$$

### 5.1 Experimental Results

We compare the efficacy of our approach described in Algorithm 3 to the method in [8], which is described in Section 3. We use PB (for PrivBasis) to denote our method, and TF (for Truncated Frequency) to denote the method in [8].

Our algorithm is adaptive based on the nature of the dataset involved and the value of $k$ desired. More specifically, the value of $\lambda$ determines how the basis set is selected. We thus roughly divide our experiments into three groups to demonstrate the efficacy of our algorithm under different scenarios.

As we point out in Section 3, the TF algorithm becomes inaccurate, and, in some cases, cannot be applied for large values of $m$. Hence, we test different values of $m$ and report the results for the value that provides the best precision.

In our experiments, we vary $\epsilon$ and report the results. We repeat all our experiments 3 times and report the mean of the results as well the standard error.

**Small $\lambda$, single basis.** When $\lambda$ is small, our PB method uses a single basis with all the top $\lambda$ frequent items. We are able to observe this scenario with the Mushroom and Pumbs_star datasets with values of $k$ less than 150. The results are shown in Figure 1 and Figure 2. The results on both datasets show that the PB method consistently and significantly out-performs the TF method both in terms of false negative rate and relative error. In fact, the performance of PB with larger $k$ significantly outperforms that of TF with a smaller $k$.

For both datasets, the FNR for PB is close to 0 even when $\epsilon$ is 0.5. In addition, the relative error is consistently small which indicates that we can get relatively high accuracy for the released itemset counts. On the other hand, the TF method has unacceptably large FNR both for larger $k$ and for smaller $\epsilon$. For example, for getting the top 100 itemsets in the Mushroom dataset, TF has FNR at over 0.6 even when $\epsilon = 1$; and for getting the top 150 itemsets in the Pumsb Star dataset, TF has FNR at over 0.7 even when $\epsilon = 1$. For obtaining the top 50 itemsets, at $\epsilon = 0.5$, TF has FNR at about 0.6 and 0.4 for the two datasets. This confirms our analysis in Section 3.

**Larger $\lambda$, small number of basis.** For larger and sparser datasets, $\lambda$ can be large enough to make the construction of a single basis unfeasible. This is the case for the retail and kosarak datasets. We run our experiments and construct bases of length 7 each as described by our algorithm in the previous section. The results for these datasets are shown in Figure 3 and Figure 4. We again see that PB out-performs TF. While the performance of PB is accurate even when $k = 400$, the PB method has acceptance FNR only for $k = 100$ and $\epsilon \geq 0.5$. An interesting observation here is that for the retail dataset, the FNR is worse than the other datasets on all accounts. Upon investigation, we realized that this is mainly due to the nature of the dataset. For larger $k$, there are many itemsets whose frequencies are lower than $f_k$ but very close $f_k$. Hence the ratio of the probability of selecting the correct top $k$ itemsets over the other is not large.

**$\lambda \approx k$, large number of basis.** For very sparse datasets, such as search log datasets, the number of frequent itemsets are largely dominated by frequent items. This is the case for the AOL dataset, for which the top 200 frequent itemsets contain 171 singletons and 29 pairs. For the TF method, it is unfeasible to run the algorithm for $m > 1$ and produce accurate results since $|\mathbf{I}|$ is quite large. The results are shown in Figure 5. This is the dataset where TF performs closest to PB, because while TF degenerates into finding all frequent singleton items, this can cover a large number of frequent itemsets. And when the problem more or less degenerates into finding the top $k$ items, the advantage of PB over TF is small.

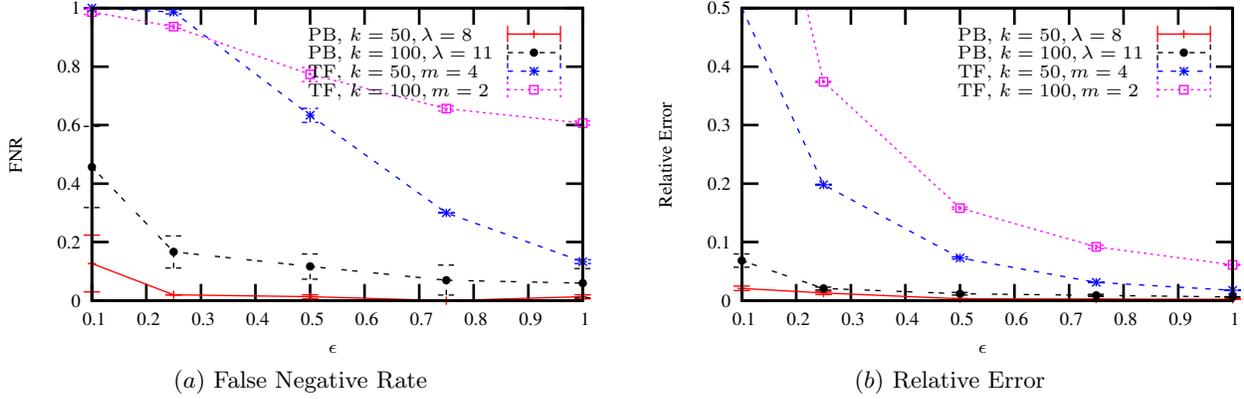

**Figure 1: Results of the PrivBasis (PB) method and the Truncated Frequency (TF) method for the Mushroom dataset, with $k = 50$ and $k = 100$; $m$ is the maximum frequent itemset length that provides the highest precision for TF.**

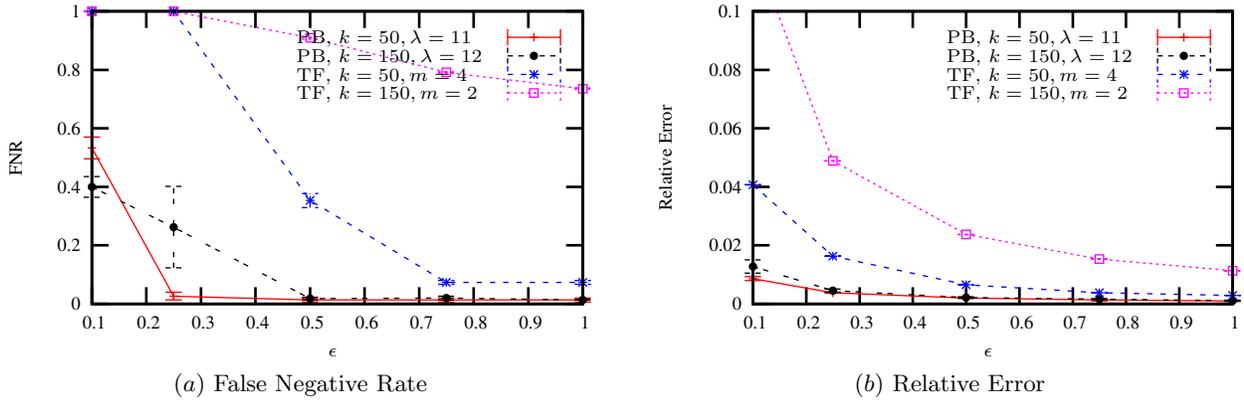

**Figure 2: Results of PB and TF for the Pumsb Star dataset, with $k = 50$ and $k = 150$; $m$ is the maximum frequent itemset length that provides the highest precision for TF.**

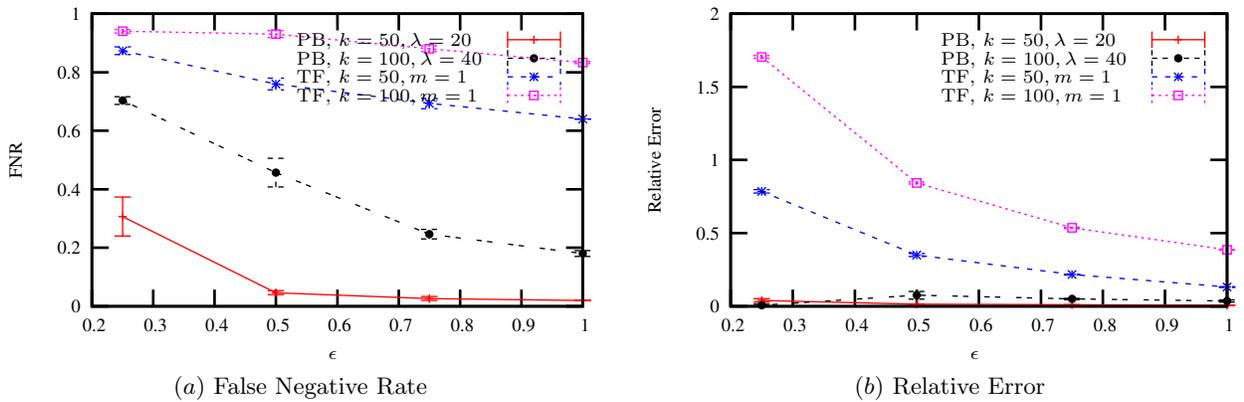

**Figure 3: Results of PB and TF for the Retail dataset, with $k = 50$ and $k = 100$; $m$ is the maximum frequent itemset length that provides the highest precision for TF.**

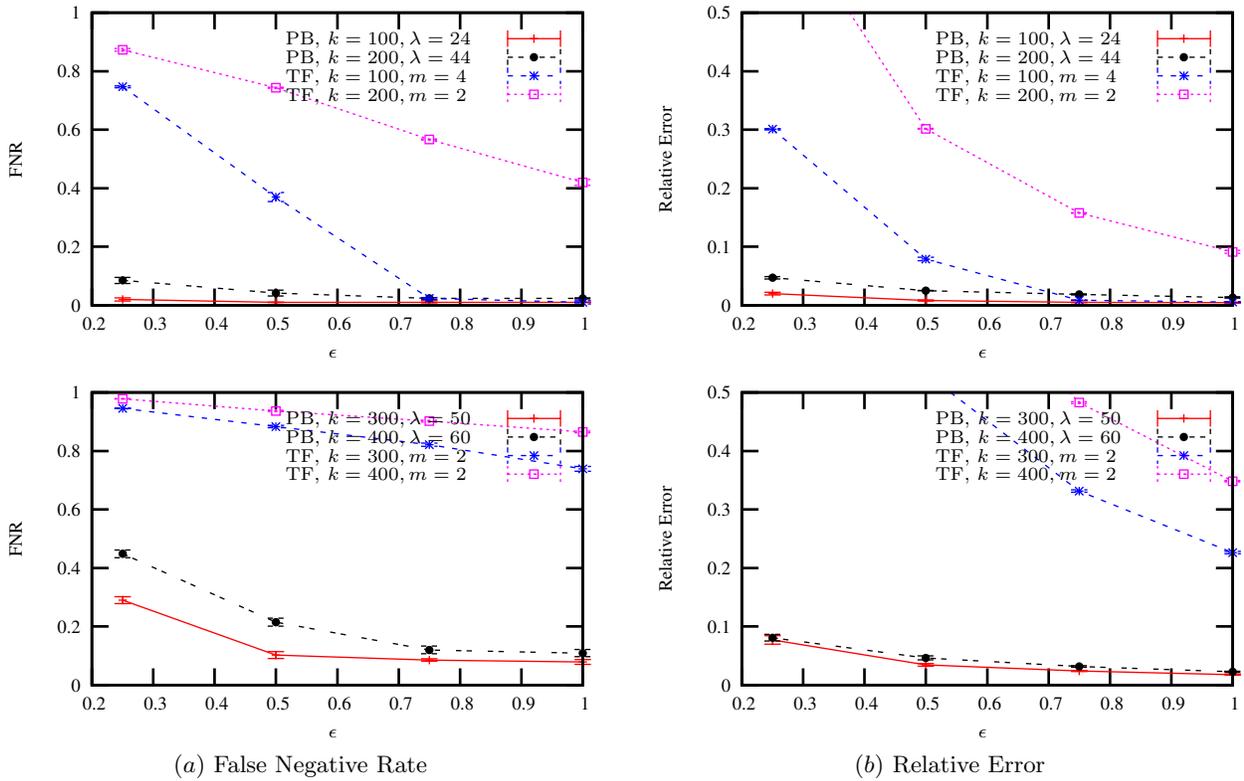

Figure 4: Results for the Kosarak dataset, first row showing results for $k = 100$ and $k = 200$; second row showing results for $k = 300$ and $k = 400$; $m$ is the maximum frequent itemset length that provides the highest precision for TF.

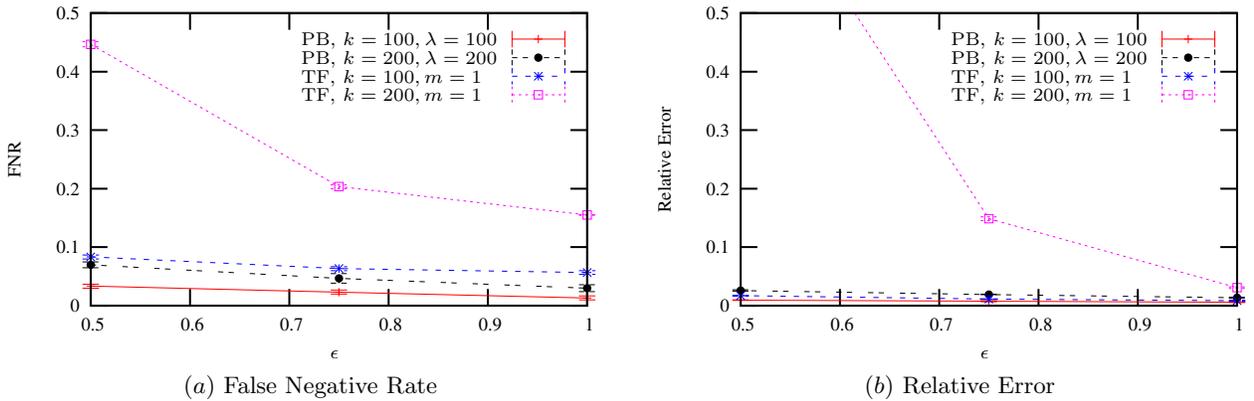

Figure 5: Results for PB and TF the AOL dataset, with $k = 100$ and $k = 200$; $m$ is the maximum frequent itemset length that provides the highest precision for TF.

## 6. RELATED WORK

Differential privacy was presented in a series of papers [15, 18, 9, 17, 16], and methods of satisfying it are presented in [16, 30, 28]. Work on differential privacy has initially focused on answering statistical queries; however, recent literature has focused on using differential privacy for data publishing scenarios [10, 19]. Differential privacy has also been employed to release contingency table [7], publish histograms [23, 33], and privately match records [25]. McSherry and Mironov [27] present differentially private recommendation algorithms in Netflix Prize competition. More recently, differential privacy has been adapted to release accurate data mining models and results [19, 29, 8].

The existing work most related to ours is Bhaskar et al. [8], which releases differentially private frequent itemsets. We have discussed this approach in detail in Section 2.

Atzori et al. [6] investigated the problem of modifying the supports of frequent itemsets, while concealing the sensitive information of individuals. It requires that each pattern derived from the released frequent itemsets have a support of either 0 or at least $k$, a positive integer threshold. This solution is based $k$-anonymity [31] privacy model, which is a much weaker privacy notion than differential privacy.

Chen et al. [14] studied the releasing of transaction dataset while satisfying differential privacy. They present an algorithm, which partitions the transaction dataset in a top-down fashion guided by a context-free taxonomy tree, and reports the noisy counts of the transactions at the leaf level. This method generates a synthetic transaction dataset, which can be then used to mine the top $k$ frequent itemsets. For the datasets we consider in this paper, this method generates either an empty synthetic dataset or a dataset that is highly inaccurate. An analysis of the method shows that this method can provide reasonable performance only when the number of items is small. (One dataset used in [14] for evaluation is the MSNBC dataset which has 17 items and about 1 million transactions.)

Work on releasing differentially private private search logs, including the AOL search log dataset, has been addressed in [26] and [21]. These works differ from our work in that they focus on releasing the top frequent keywords that occur in the search logs, and does not release any information about how frequent itemsets with size 2 or higher. This is essentially mining for frequent itemsets of length 1. Another difference is that their approach assume that the keywords in the dataset are not public knowledge, whereas we assume $\mathbf{I}$ is public. As a result, their approach satisfies a relaxed version of differential privacy similar to the notion of $(\epsilon, \delta)$-differential privacy.

In addition, there is another series of works [32, 24, 20, 34, 13, 14] on publishing anonymized transaction data, instead of releasing privacy-preserving mining results [19, 29, 8, 6]. Terrovitis et al. [32] apply a relaxation of $k$-anonymity on transaction dataset, by requiring that for each itemset with the length of at most $m$, the number of transactions in the dataset containing this itemset is either 0 or at least $k$. He and Naughton [24] enhance [32] by strictly imposing $k$-anonymity. The two solutions [32, 24] treat each item in the dataset equally. Different from them, the schemes [20, 34] divide items into sensitive and non-sensitive ones, and assume that an adversary can only get the background knowledge about the non-sensitive items. The algorithms in [20, 34] ensure that the inference from non-sensitive items to a sensitive one is lower than a threshold. Cao et al. [13] relax the assumption, and allow an attacker to include sensitive items in his/her background knowledge. They provide a privacy principle $\rho$-uncertainty, which postulates that the confidence of inferring a sensitive item from any itemset (consisting of both sensitive and non-sensitive items) be lower than $\rho$, a threshold.

## 7. CONCLUSION

In this paper, we have introduced PrivBasis, a novel method of publishing frequent itemsets with differential privacy guarantees. The intuition behind PrivBasis is simple. Given some minimum support threshold, $\theta$, one can construct a basis set $\mathbf{B} = \{B_1, B_2, \ldots, B_w\}$ such that any itemset with frequency higher than $\theta$ is a subset of some basis $B_i$. We have introduced techniques for privately constructing basis sets, and for privately reconstructing the frequencies of all subsets of $B_i$'s with reasonable accuracy. One can then select the most frequent itemsets from such reconstructed subsets. We have conducted experiments on 5 real datasets commonly used for frequent itemset mining purposes, and the results show that our approach greatly outperforms the current state of the art. Our approach can be viewed as a dimension reduction to deal with the curse of dimensionality in private data analysis and data anonymization.

## 8. ACKNOWLEDGEMENTS

The work reported in this paper was partially supported by the Air Force Office of Scientific Research MURI Grant FA9550-08-1-0265, and by the National Science Foundation under Grant No. 1116991.